\pdfoutput=1

\documentclass[11pt]{article}

\usepackage{acl}

\usepackage{times}
\usepackage{latexsym}
\usepackage{soul}
\usepackage[T1]{fontenc}
\usepackage{verbatim}
\usepackage{booktabs}
\usepackage{hyperref}
\usepackage{multirow}
\usepackage[table]{colortbl}
\usepackage{graphicx}
\usepackage{amsmath}
\usepackage{subcaption}
\usepackage[most]{tcolorbox}

\usepackage[utf8]{inputenc}

\usepackage{microtype}

\usepackage{inconsolata}
\usepackage{graphicx}
\usepackage{subcaption}

\title{\textit{Heterogeneity over Homogeneity}: Investigating Multilingual Speech Pre-Trained Models for Detecting Audio Deepfake}

\author{
  Orchid Chetia Phukan$^{1\dagger\scriptsize\thanks{\hspace{6pt}Corresponding Author}}$ , 
  Gautam Siddharth Kashyap$^{1}$\thanks{\hspace{6pt}Authors contributed equally as first authors} , 
  Arun Balaji Buduru$^{1}$\\ 
  {\bf Rajesh Sharma}$^{1,2}$\\
  \textsuperscript{1}IIIT-Delhi, India\\ 
  \textsuperscript{2}University of Tartu, Estonia\\ 
  \texttt{orchidp@iiitd.ac.in}
}

\begin{document}
\maketitle
\begin{abstract}

In this work, we investigate multilingual speech Pre-Trained models (PTMs) for Audio deepfake detection (ADD). We hypothesize that multilingual PTMs trained on large-scale diverse multilingual data gain knowledge about diverse pitches, accents, and tones, during their pre-training phase and making them more robust to variations. As a result, they will be more effective for detecting audio deepfakes. To validate our hypothesis, we extract representations from state-of-the-art (SOTA) PTMs including monolingual, multilingual as well as PTMs trained for speaker and emotion recognition, and evaluated them on ASVSpoof 2019 (ASV), In-the-Wild (ITW), and DECRO benchmark databases. We show that representations from multilingual PTMs, with simple downstream networks, attain the best performance for ADD compared to other PTM representations, which validates our hypothesis. 
We also explore the possibility of fusion of selected PTM representations for further improvements in ADD, and we propose a framework, \textbf{MiO} (\textbf{M}erge \textbf{i}nto \textbf{O}ne) for this purpose. With \textbf{MiO}, we achieve SOTA performance on ASV and ITW and comparable performance on DECRO with current SOTA works.

\end{abstract}

\section{Introduction}

The popularity of audio deepfakes has raised multiple concerns in areas dealing with personal and public security due to its capability to impersonate and share false, often malicious information. Scammers, for example, have utilized audio deepfake to mimic a German executive, successfully convincing a transfer of 220,000 Euros to a Hungarian supplier \cite{stupp2019fraudsters}. Thus, checking and evaluating authenticity of any audio content is important through robust and reliable measures. Motivated by this, in this work,  we focus on Audio Deepfake Detection (ADD). \par 

To combat this progressing issue, various detection methods have been proposed \cite{hanilci15_interspeech, qian2016deep, ma2021improved, luo2021capsule}. These works leveraged statistical attributes or the raw audio as input to the models. However, with the wide-scale accessibility of Pre-Trained Models (PTMs), ADD as a task has undergone exponential advancement. Representations from PTMs are used as input features to downstream ADD and it comes with a series of benefits, which include higher accuracy in ADD and saving time as well as resources, in building ADD systems from the ground up. PTMs come in various architectures as well as varied pre-training schemes and are trained on large-scale datasets. They can be either trained in a supervised (Eg. Whisper \cite{radford2023robust}) or in a self-supervised manner (Eg. Wav2vec2 \cite{baevski2020wav2vec}) and also on either single or multiple languages. Despite these PTMs being pre-trained on only real speech data, representations from these PTMs have shown exceptional performance in identifying real content from their fake counterparts \cite{yang21c_interspeech}. 

Our work relies on the hypothesis that \textit{multilingual PTMs trained on large-scale diverse multilingual data, acquire knowledge about diverse pitches, accents, tones, and are more robust to variations, hence will be more effective for identifying audio deepfakes than other PTMs}. So to validate our hypothesis, we extract representations from eight state-of-the-art (SOTA) PTMs including multilingual (XLS-R, Whisper, MMS), monolingual (Unispeech-SAT, WavLM, Wav2vec2), speaker recognition (x-vector), and emotion recognition (XLSR\_emo) and evaluate them with two simple probing networks (Fully Connected Network (FCN), Convolution Neural Network (CNN)) on three benchmark datasets ASVSpoof 2019 (ASV), In-the-Wild (ITW), and DECRO. We also investigate by combining representations from different PTMs as it has been seen in other speech processing tasks such as speech recognition \cite{arunkumar22b_interspeech} that certain representations act as complementary to each other and we propose a framework, \textbf{M}erge \textbf{i}nto \textbf{O}ne (\textbf{MiO}) for the same. To the best of our knowledge, this is the first study, to explore fusion of PTM representations for ADD. Our study makes the following contributions: \par

\begin{itemize}
    \item Comprehensive empirical study to demonstrate the performance of multilingual PTMs for ADD, which have shown top performance in comparison to its other PTM counterparts across the three datasets.
    \item A novel approach to fuse representations from different PTMs, namely \textbf{MiO}. Our approach shows demonstrable improvement in performances over individual representations. It achieves SOTA in terms of Equal Error Rate (EER) in ASV, ITW, and competitive performance in DECRO.
\end{itemize}


\section{Related Works}
\label{sec:Related Works}
In this section, we give an overview of various prolific ADD methods proposed. ADD as a task has caught the attention with the release of the ASVspoof 2015 \cite{wu2015asvspoof} database. Initially, researchers built GMM and SVM-based modeling approaches with statistical audio features as input \cite{sahidullah15_interspeech}. 
Previous works have also harnessed neural network-based models such as CNN, RNN, etc. for ADD \cite{tom18_interspeech, gomezalanis19_interspeech, alzantot19_interspeech}.  

Researchers have exploited self-supervised learning (SSL) modeling approaches for ADD \cite{lee2023experimental, shim20_interspeech, jiang20b_interspeech}. Further, different types of PTMs such as Wav2vec, HuBERT, TERA, Mockingjay, etc also been explored for ADD \cite{eom2022anti, yang21c_interspeech}. \citealt{wang23x_interspeech} showed that generalization of ADD systems increases with combination of Wav2vec, prosodic, and pronounciation information as input features. In this work, we evaluate eight PTMs to validate our hypothesis that multilingual PTMs trained on extensive and diverse multilingual datasets allow them to capture knowledge related to diverse pitch, accent, tone, and so on. This broad exposure enhances their robustness to different variations in audio signals. As a result, the representations learned by these PTMs are particularly effective for discerning audio deepfakes when compared to representations from other PTMs.

\section{Pre-Trained Models}

We compile the top-performing PTMs for our experiments. For multilingual PTMs we choose, \textbf{XLS-R} \cite{babu22_interspeech}, \textbf{Whisper} \cite{radford2023robust}, and \textbf{Massively Multilingual Speech (MMS)} \cite{pratap2023scaling}. XLS-R was pre-trained on 128 languages while Whisper on 96 and MMS over 1400 languages. Whisper improves XLS-R in various downstream speech processing tasks while MMS improves over Whisper. We selected the monolingual PTMs (\textbf{WavLM}, \textbf{Unispeech-SAT}, \textbf{Wav2vec2}) based on the SUPERB \cite{yang21c_interspeech}. WavLM and Unispeech-SAT have shown SOTA performance on SUPERB so we choose them. Wav2vec2 \cite{baevski2020wav2vec} has not shown top performance like WavLM \cite{chen2022wavlm} and Unispeech-SAT \cite{chen2022unispeech} on SUPERB, however, as previous works have shown its efficacy for ADD \cite{zhang2023audio, cai2023waveform}, so we selected it. \par

 Additionally, models pre-trained for more specific tasks such as speaker recognition PTM \cite{ma2023boost} and models trained for emotion recognition \cite{conti2022deepfake} show exceptional performance for ADD, so we included them in our experiments. As speaker recognition PTM, we consider, \textbf{x-vector} \cite{snyder2018x} and as emotion recognition PTM we use, \textbf{XLSR\_emo} \cite{cahyawijaya23_interspeech}. Additional details regarding these selected PTMs are available in Appendix 9.2. 

\section{Modeling}
\label{Downstream Model}

\begin{figure}[htbp]
\centering
    \begin{subfigure}{0.08\textwidth}
        \centering
        \includegraphics[width=1.2\textwidth]{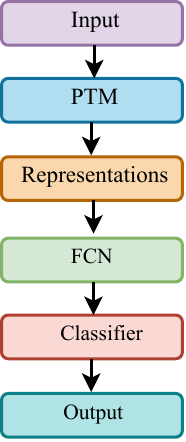} 
        \caption{FCN}
        \label{fc}
    \end{subfigure}%
    \hspace{0.05\textwidth}
    \begin{subfigure}{0.105\textwidth}
        \centering
        \includegraphics[width=0.95\textwidth]{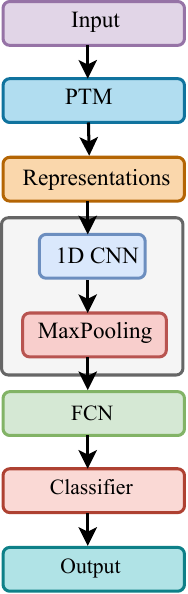} 
        \caption{CNN}
        \label{1dcnn}
    \end{subfigure}%
    \caption{Modeling Approaches}
    \label{modelarchi}
\end{figure}

As we are evaluating how representations of different PTMs will behave for ADD, we keep the PTM layers frozen and keep the downstream modeling as simple as possible. We experimented with two modeling approaches (see Figure \ref{fc} and \ref{1dcnn}). For the first approach (FCN), we employ an FCN on the extracted PTM representations and for the second (CNN), we use a 1D-CNN layer on top of representations followed by a Maxpooling layer and FCN. Softmax is used as the activation function in the output layer of the models which gives output as probabilities. \par

\begin{figure}[hbt!]    
\centering
      \includegraphics[width=0.3\textwidth, height=0.4\textwidth]{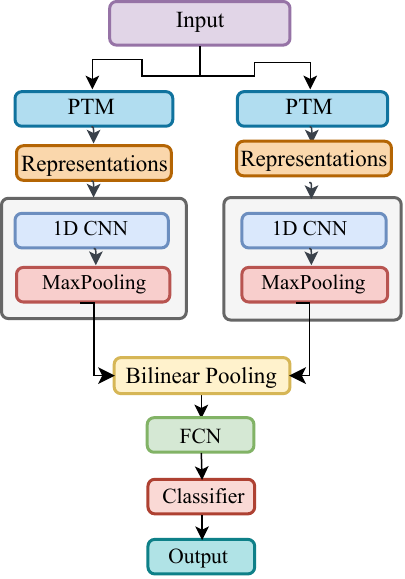}
      \caption{Proposed Modeling Framework for Fusion of PTM Representations, \textbf{MiO}}
        \label{fusion}     
\end{figure} 

\noindent\textbf{Merge into One}: For fusing representations of different PTMs, we propose \textbf{MiO}. The model architecture is shown in Figure \ref{fusion}. Here, we follow the same modeling pattern for each incoming representation as the second approach mentioned above. Then we apply linear projection to a dimension of size 120 followed by bilinear pooling (BP), which allows effective interaction between the features as shown by \cite{kumar2022hate}. BP is the outer product of two vectors p and q of dimension (D,1) and (D,1) such that the resultant will be a matrix of dimension (D, D) and it is given as: 
\begin{equation}
\mathbf{BP_{D,D}} = \mathbf{p_{D,1}} \otimes \mathbf{q_{D,1}} = \mathbf{p} \mathbf{q}^T
\label{eq:BP}
\end{equation}
Linear Projection to lower dimension is carried out for computational resource constraints as the resultant of BP results in a matrix of much bigger shape. The resultant of BP is flattened and passed through an FCN.\par 

\begin{table}[ht]
\scriptsize
  \centering
  \begin{tabular}{l|c|c|c|c}
    \toprule
    \textbf{PTM} & \textbf{ASV} & \textbf{ITW} & \textbf{D-C} & \textbf{D-E} \\
    \midrule
        XLS-R & \cellcolor{blue!25}\textbf{1.67} & \cellcolor{blue!25}\textbf{0.24} & \cellcolor{green!25} \textbf{1.42} & \cellcolor{yellow!25} \textbf{0.12} \\
        Whisper & \cellcolor{yellow!25} \textbf{2.34} & \cellcolor{green!25} \textbf{0.73} & \cellcolor{blue!25}\textbf{0.69} & \cellcolor{blue!25}\textbf{0.11} \\
        MMS & \cellcolor{green!25} \textbf{3.10} & \cellcolor{yellow!25}\textbf{0.31} & \cellcolor{yellow!25} \textbf{1.11} & \cellcolor{green!25} \textbf{0.19} \\
        Unispeech-SAT & 9.97 & 2.36 & 2.14 & 0.54 \\
        WavLM (Base) & 10.46 & 8.48 & 4.22 & 0.57 \\
        WavLM (Large) & 10.23 & 8.41 & 4.20 & 0.60 \\
        Wav2Vec2 & 12.45 & 16.54 & 6.95 & 1.10 \\
        x-vector & 9.49 & 0.98 & 2.65 & 0.66 \\
        XLSR\_emo & 9.36 & 2.70 & 3.11 & 0.18 \\
        \hline
    \end{tabular}
\caption{EER (\%) scores for FCN models with different PTM representations; D-C, D-E represents Chinese and English Set of DECRO respectively}
\label{fcnsingle}
\end{table}

\begin{table}[ht]
\scriptsize
  \centering
  \begin{tabular}{l|c|c|c|c}
    \toprule
    \textbf{PTM} & \textbf{ASV} & \textbf{ITW} & \textbf{D-C} & \textbf{D-E} \\
    \midrule
        XLS-R & \cellcolor{blue!25}\textbf{1.03} & \cellcolor{blue!25}\textbf{0.17} & \cellcolor{green!25} \textbf{1.42} & \cellcolor{green!25} \textbf{0.07} \\
        Whisper & \cellcolor{green!25} \textbf{2.02} & \cellcolor{green!25} \textbf{0.22} & \cellcolor{blue!25}\textbf{0.58} & \cellcolor{blue!25}\textbf{0.06} \\
        MMS & \cellcolor{yellow!25}\textbf{1.50} & \cellcolor{yellow!25}\textbf{0.20} & \cellcolor{yellow!25}\textbf{0.70} & \cellcolor{yellow!25}\textbf{0.09} \\
        Unispeech-SAT & 9.76 & 2.20 & 1.89 & 0.42 \\
        WavLM (Base) & 9.90 & 8.31 & 4.21 & 0.44 \\
        WavLM (Large) & 9.30 & 7.60 & 3.14 & 0.45 \\
        Wav2Vec2 & 10.33 & 14.77 & 6.66 & 1.05 \\
        x-vector & 8.61 & 0.27 & 2.62 & 0.92 \\
        XLSR\_emo & 9.20 & 1.57 & 2.83 & 0.10 \\
        \hline
    \end{tabular}
\caption{EER (\%) scores for CNN models with different PTM representations; D-C, D-E represents Chinese and English Set of DECRO respectively}
\label{cnnsingle}
\end{table}

We keep the number of epochs at 20 and the batch size of 32 for all the modeling approaches. We use Cross-entropy as the loss function and Adam as the optimizer. We use \textit{Tensorflow} library for our experiments. For reproducing our experiments, we will make our codebase available here\footnote{\url{https://github.com/orchidchetiaphukan/MultilingualPTM_ADD_NAACL24}}.

\section{Experiments}
\label{sec:Experiments}

\subsection{Benchmark Datasets}
\label{ssec:Benchmark Datasets}
We selected three benchmark datasets for our experiments. They are \textbf{ASVSpoof 2019 (ASV)}
\cite{wang2020asvspoof}, \textbf{In-the-Wild Dataset (ITW)} \cite{muller22_interspeech}, and \textbf{DECRO} \cite{ba2023transferring}. Details regarding the datasets, data preprocessing, and experimental setting is provided in Appendix 9.1.

\subsection{Experimental Results}
\label{ssec:Experimental Results}
For ASV and DECRO, the results presented are on the official testing split, and for ITW, the scores are the average across five splitting seeds as no official split was given. On DECRO, we built and trained models individually on the Chinese and English set of DECRO. We use D-C and D-E as notations for DECRO Chinese and English sets respectively.\par

Table \ref{fcnsingle} and \ref{cnnsingle} presents the EER scores for FCN and CNN models for different PTM representations as input features. Models trained on multilingual PTMs (XLS-R, Whisper, MMS) representations performed the best with lowest EERs in comparison with other PTMs. Within the multilingual PTMs, XLS-R achieves the lowest EERs in ASV and ITW while Whisper representations report the lowest in DECRO. MMS showed mixed performance achieving the second lowest EER in ITW and D-C with FCN, while with CNN it got the second least EER in ASV, ITW, and D-C. \textit{This validates our hypothesis that multilingual PTMs will be more effective for ADD} due to their pre-training on extensive and varied multilingual datasets. As a result, they acquire information on a wide range of pitch, accents, and tones during the pre-training phase. This acquisition of diverse knowledge enhances their ability to effectively recognize and identify variations.  Overall, CNN models showed superior performance to FCN models due to their ability to capture further important features.

\begin{figure*}[htbp]
    \begin{subfigure}{0.20\textwidth}
        \centering
        \includegraphics[width=\linewidth]{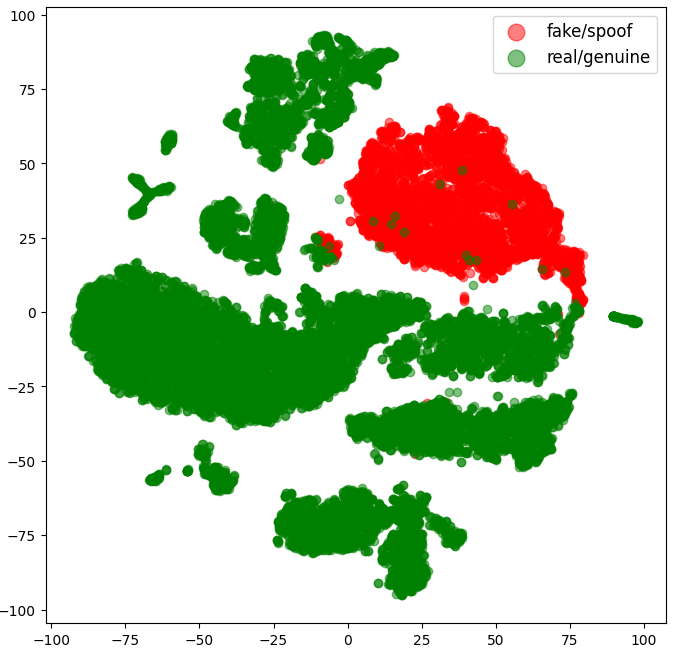}
        \caption{XLS-R}
    \end{subfigure}%
    \begin{subfigure}{0.2\textwidth}
        \centering
        \includegraphics[width=\linewidth]{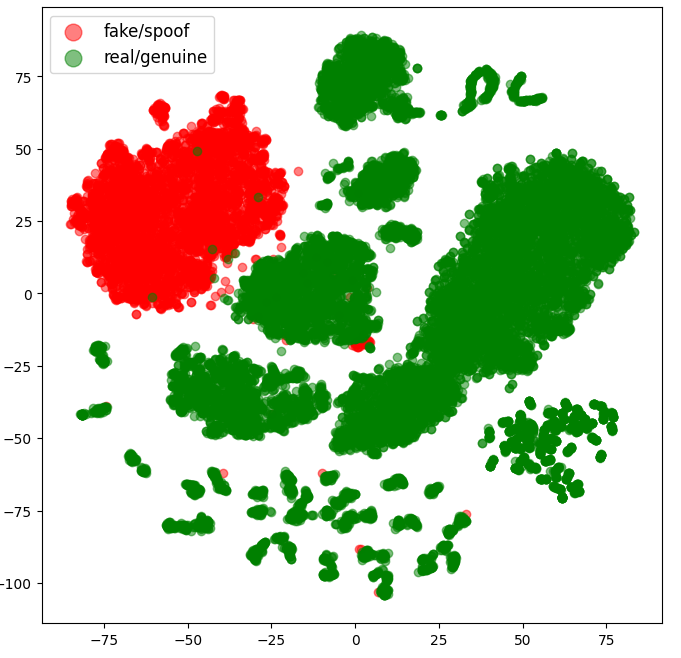}
        \caption{Whisper}
    \end{subfigure}%
    \begin{subfigure}{0.2\textwidth}
        \centering
        \includegraphics[width=\linewidth]{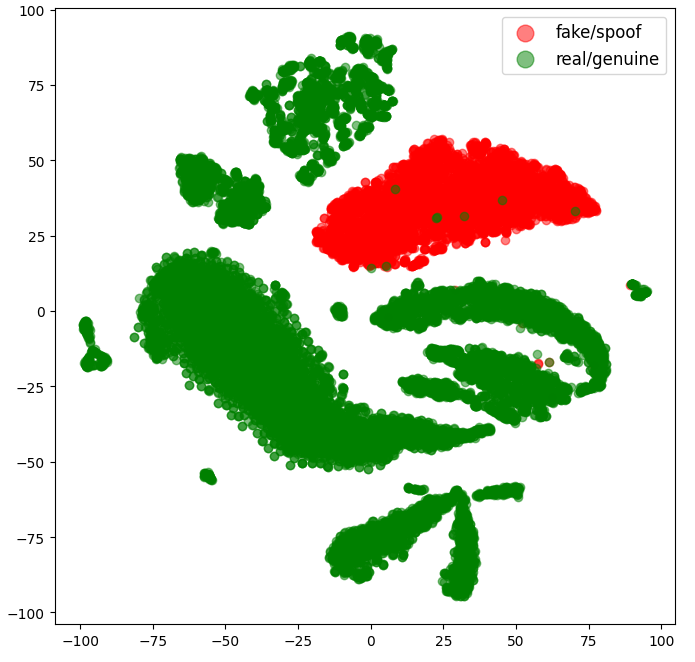}
        \caption{MMS}
    \end{subfigure}%
    \begin{subfigure}{0.2\textwidth}
        \centering
        \includegraphics[width=\linewidth]{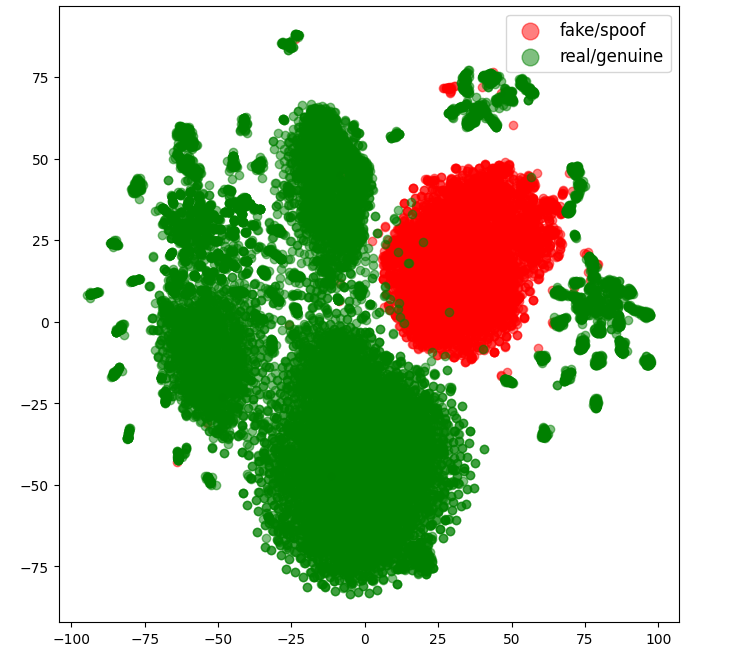}
        \caption{Unispeech-SAT}
    \end{subfigure}%
    \begin{subfigure}{0.208\textwidth}
        \centering
        \includegraphics[width=\linewidth]{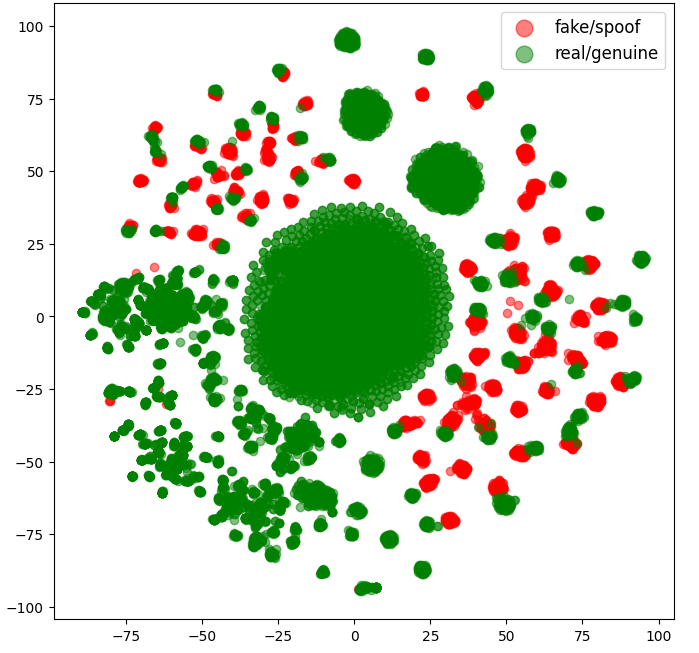}
        \caption{x-vector}
    \end{subfigure}%
    \caption{t-SNE plots of different PTM representations on D-E}
    \label{tsnede}
\end{figure*}

We experimented with both WavLM (Base) and WavLM (Large) as WavLM (Large) holds the top position in SUPERB and also to see if the version with more parameters comes with the benefit of increased ADD performance. WavLM (Large) performs better than WavLM (Base) in some instances which might be due to its larger size. However, Unispeech-SAT has superior performance compared to WavLM (Large) while having the same number of parameters as WavLM (Base) and the best among the monolingual PTMs, Unispeech-SAT achieved the lowest EER in most instances with both FCN (9.97\%, 2.36\%, 2.14\%, 0.54\% in ASV, ITW, D-C, D-E respectively)  and CNN (2.2\%, 1.89\%, 0.42\% in ITW, D-C, D-E respectively). This can be attributed to speaker-aware pre-training of Unispeech-SAT, which leads to capturing various speech attributes such as pitch, accent, etc, more effectively and that helps in identifying deepfakes with more efficacy than its other monolingual counterparts. We can also see that representations of x-vector and XLSR\_emo are performing better in some instances than the monolingual PTMs as they are trained on more specific non-semantic tasks leading to capture speech attributes far better than the monolingual PTMs for ADD. Wav2vec2 performed the worst among all the PTMs considered in our study showing its ineffectiveness in capturing attributes important for segregrating fake from real audio. Visualization of the representational space from the PTMs last hidden state are shown in Appendix (See Figure \ref{tsneasv}, \ref{tsneitw}, \ref{tsnedc}, \ref{tsnede} for ASV, ITW, D-C, and D-E respectively). We observe better clustering across the classes (real/fake) for representations from multilingual PTMs.\par

\begin{table}[ht]
\scriptsize
  \centering
  \begin{tabular}{l|c|c|c|c}
    \toprule
    \textbf{PTM Combinations} & \textbf{ASV} & \textbf{ITW} & \textbf{D-C} & \textbf{D-E} \\
    \midrule
    XLS-R + Whisper &0.95 &0.27 &1.08 & \cellcolor{blue!25}\textbf{0.05}\\
    XLS-R + MMS & 0.56 & 0.29 &1.62 & \cellcolor{yellow!25} \textbf{0.06} \\
    XLS-R + Unispeech-SAT & \cellcolor{yellow!25} \textbf{0.45} & \cellcolor{yellow!25} \textbf{0.11} &1.03 &0.13\\
    XLS-R + WavLM (Base) &0.82 &0.16 &1.36 &0.14\\
    XLS-R + WavLM (Large) &0.72 &0.14 &1.16 &0.12\\
    XLS-R + Wav2Vec2 &1.06 & \cellcolor{green!25} \textbf{0.12} &1.80 &0.11\\
    XLS-R + x-vector &\cellcolor{blue!25}\textbf{0.41} &\cellcolor{blue!25}\textbf{0.07} &1.63 &0.46\\
    XLS-R + XLSR\_emo &1.35 &0.21 &1.60 &0.12\\
    Whisper + MMS &2.24 &0.15 & \cellcolor{green!25} \textbf{0.27} & \cellcolor{green!25} \textbf{0.08}\\
    Whisper + Unispeech-SAT &2.16 &1.03 & 0.46 &0.28\\
    Whisper + WavLM (Base) &1.90 &0.97 &2.95 &0.15\\
    Whisper + WavLM (Large) &1.95 &0.91 &2.10 &0.13\\
    Whisper + Wav2Vec2 &2.19 &1.08 &0.98 &0.65\\
    Whisper + x-vector &3.30 &0.22 &0.91 &0.32\\
    Whisper + XLSR\_emo &1.81 &0.63 &0.88 &0.21\\
    MMS + Unispeech-SAT &4.44 &0.17 & \cellcolor{yellow!25}\textbf{0.19} &0.24 \\
    MMS + WavLM (Base) &0.99 &3.50 & 0.21 &0.25 \\
    MMS + WavLM (Large) &1.00 & 3.10 & \cellcolor{blue!25} \textbf{0.15} &0.22 \\
    MMS + Wav2Vec2 & \cellcolor{green!25} \textbf{0.50} &0.22 &0.39 &0.33 \\
    MMS + x-vector &5.40 &0.14 &0.77 &0.25 \\
    MMS + XLSR\_emo &1.80 &0.36 &0.81 &0.32 \\
    Unispeech-SAT + WavLM (Base) &10.18 &2.79 &2.31 &0.48\\
    Unispeech-SAT + WavLM (Large) &9.19 &2.99 &2.11 &0.41\\
    Unispeech-SAT + Wav2Vec2 &9.74 &2.55 &2.88 &0.59\\
    Unispeech-SAT + x-vector &5.82 &0.15 &2.56 &0.54\\
    Unispeech-SAT + XLSR\_emo &6.80 &1.70 &2.09 &0.57\\
    WavLM (Base) + Wav2Vec2 &12.46 &8.54 &1.93 &0.51\\
    WavLM (Base) + x-vector &6.03 &0.19 &3.04 &0.61\\
    WavLM (Base) + XLSR\_emo &7.91 &2.31 &2.64 &0.21\\
    WavLM (Large) + Wav2Vec2 &11.36 &7.14 &1.44 &0.54\\
    WavLM (Large) + x-vector &5.01 &0.19 &2.21 &0.60\\
    WavLM (Large) + XLSR\_emo &6.92 &2.20 &2.01 &0.18\\
    Wav2Vec2 + x-vector &7.31 &0.26 &3.51 &0.47\\
    Wav2Vec2 + XLSR\_emo &7.50 &1.91 &2.87 &0.32\\
    x-vector + XLSR\_emo &7.89 &0.43 &1.11 &0.71\\
    
    \hline 
  \end{tabular}
  \caption{EER (\%) scores for different PTM representations combinations with \textbf{MiO}}
  \label{fusion_eer}
\end{table}

Table \ref{fusion_eer} shows the EER scores with combined representations of different PTMs. With the fusion of XLS-R and x-vector representations, we got the lowest EER score in ASV and ITW which shows that combining speaker-specific informative features leads to further gain in performance and these representations are acting as complementary to each other. In D-C, the fusion of MMS and WavLM (Large), and in D-E, XLS-R and Whisper pair, reported the lowest EER, which shows that these multilingual PTM's representations are showing additive behavior, leading to further lowering of EER. However, in some, instances the fusion of certain PTM representations leads to degradation of performance compared to its individual performance, such as the combination of XLS-R and XLSR\_emo gave 1.35\% and 0.21\% EER in ASV, ITW respectively which is lower than individual EER of XLSR\_emo, but higher than XLS-R (1.03\% in ASV). This can be depicted as contradictory behavior shown by the representations. As additional experiments, we carried out a cross-corpus evaluation (see Tables \ref{120_cross} and \ref{240_cross} in Appendix). We found that models trained on multilingual PTM representations, generalize better in cross-corpus evaluation.\par

\begin{table}[htbp]
\scriptsize
\centering
\begin{tabular}{l|l|c}
\toprule
Dataset & Model & EER (\%) \\
\midrule
ASV & CQT-DCT-LCNN \cite{lavrentyeva2019stc} & \cellcolor{yellow!25}\textbf{1.84} \\
    & \textbf{MiO(XLS-R + x-vector)} &  \cellcolor{blue!25}\textbf{0.41} \\
\midrule
ITW & STATNet \cite{ranjan2022statnet} & \cellcolor{yellow!25}\textbf{0.20} \\
    & \textbf{MiO(XLS-R + x-vector)} &  \cellcolor{blue!25}\textbf{0.07}\\
    \midrule
D-E & Res-TSSDNet \cite{ba2023transferring} &  \cellcolor{blue!25}\textbf{0.02}\\
    & \textbf{MiO(XLS-R + Whisper)} & \cellcolor{yellow!25}\textbf{0.04}\\
\bottomrule
\end{tabular}
\caption{Comparison with SOTA on ASV, ITW, and D-E in terms of EER(\%); \textbf{MiO(XLS-R + x-vector)}, \textbf{MiO(XLS-R + Whisper)} represents the proposed methodology \textbf{MiO} with combination of XLS-R, x-vector and XLS-R, Whisper representations}
\label{eer_comparison_table}
\end{table}

\subsection{Comparison with State-of-the-art}
\label{comp_sota}

We compare the proposed approach, \textbf{MiO} with previous SOTA works on respective datasets. Table \ref{eer_comparison_table}, presents the comparison with SOTA studies on ASV, ITW, and D-E respectively. D-C was used as a testing set for evaluating the transferability of ADD systems from English to Chinese by \citealt{ba2023transferring}, so previous works trained on D-C and evaluated on D-C are not present. In ASV and ITW, we report the lowest EER compared to existing SOTA works, and for D-E, we report competitive performance in comparison to existing SOTA work.

\section{Conclusion}
\label{sec:Conclusions}
In this work, we validated our hypothesis that multilingual PTMs pre-trained on large diverse multilingual data will be more effective for ADD as they learn diverse pitches, accents, and tones during their pretraining phase and are more robust to variations. We carried out a comprehensive empirical analysis by extracting representations from eight PTMs and our findings show that representations from multilingual showed the lowest EER on three benchmark datasets ASV, ITW, and DECRO. Also, we found that fusion of representations from PTMs lead to a further drop in EER and for this, we proposed, \textbf{MiO}. We report SOTA performance in ASV, ITW and competitive performance in DECRO in comparison to previous SOTA works with \textbf{MiO}.

\section{Limitations}
We have considered only eight PTMs and this may limit our findings, so in the future, we will consider more relevant PTMs. Also, results varies with different downstream networks as shown by \cite{zaiem23b_interspeech}  and we only experimented with two downstream networks. So, we will extend this by evaluating more downstream networks. Also, we will also look into why certain PTMs representations combinations works better than others. 

\section{Ethics Statement}

Deepfakes' have a significant impact on the privacy and integrity of individuals. It is important to address the ethical implications of research conducted on Deepfakes. This work ensures that privacy and integrity of specific individuals or organizations are not revealed and is not affected. The data used for research in this work is collected from publicly available datasets and are anonymized. The experimental results and interpretations also do not have any ethical implications.

\bibliography{main}

\section{Appendix}
\subsection{Dataset}

Additional information regarding the benchmark datasets considered in our study is given as follows: \newline
\textbf{ASVSpoof 2019\footnote{\url{https://www.asvspoof.org/index2019.html}}}
: Voice Cloning Toolkit (VCTK), a multispeaker English corpus is used as a base database. It contains audio clips from 107 (46 male, 61 female) speakers. Various spoofing algorithms were employed to create counterfeit versions of the authentic clips. These algorithms include SOTA text-to-speech synthesis techniques as well as different voice conversion methods. The availability of large-scale labeled audio recordings makes ASV a valuable resource for training and testing ML models to identify fake audio. We use the Logical Access (LA) database from ASV. We train, validate, and evaluate the models on the official split given by \citealt{wang2020asvspoof}.

\noindent\textbf{In-the-Wild Dataset\footnote{\url{https://deepfake-demo.aisec.fraunhofer.de/in_the_wild}}}:  This dataset comprises 37.9 hours of audio content and features English-speaking celebrities and politicians. It encompasses fake audio encountered in various real-life scenarios, presenting a challenge in distinguishing it from genuine recordings. The clips associated with the fake audio associated with a particular celebrity/politician are collected from openly available social media sites and video-sharing platforms. This dataset acts as an important resource for evaluating ADD systems on real-world data. For ITW, there is no official split given so we split the dataset as 70\% as training, 10\% as validation, and 20\% as test set.\par

\noindent\textbf{DECRO\footnote{\url{https://zenodo.org/records/7603208}}}: ADD models trained on one language fail when evaluated in zero-shot format in some other language. So as to make up for this, \citealt{ba2023transferring} proposed the DECRO dataset for the evaluation of ADD systems in a cross-lingual manner. It comprises fake and real audio clips in English (DECRO-E) and Chinese (DECRO-C). We use the official split given by \citealt{ba2023transferring} for training, validation, and evaluation. \par

\begin{figure}[htbp]
    \includegraphics[width=1\linewidth]{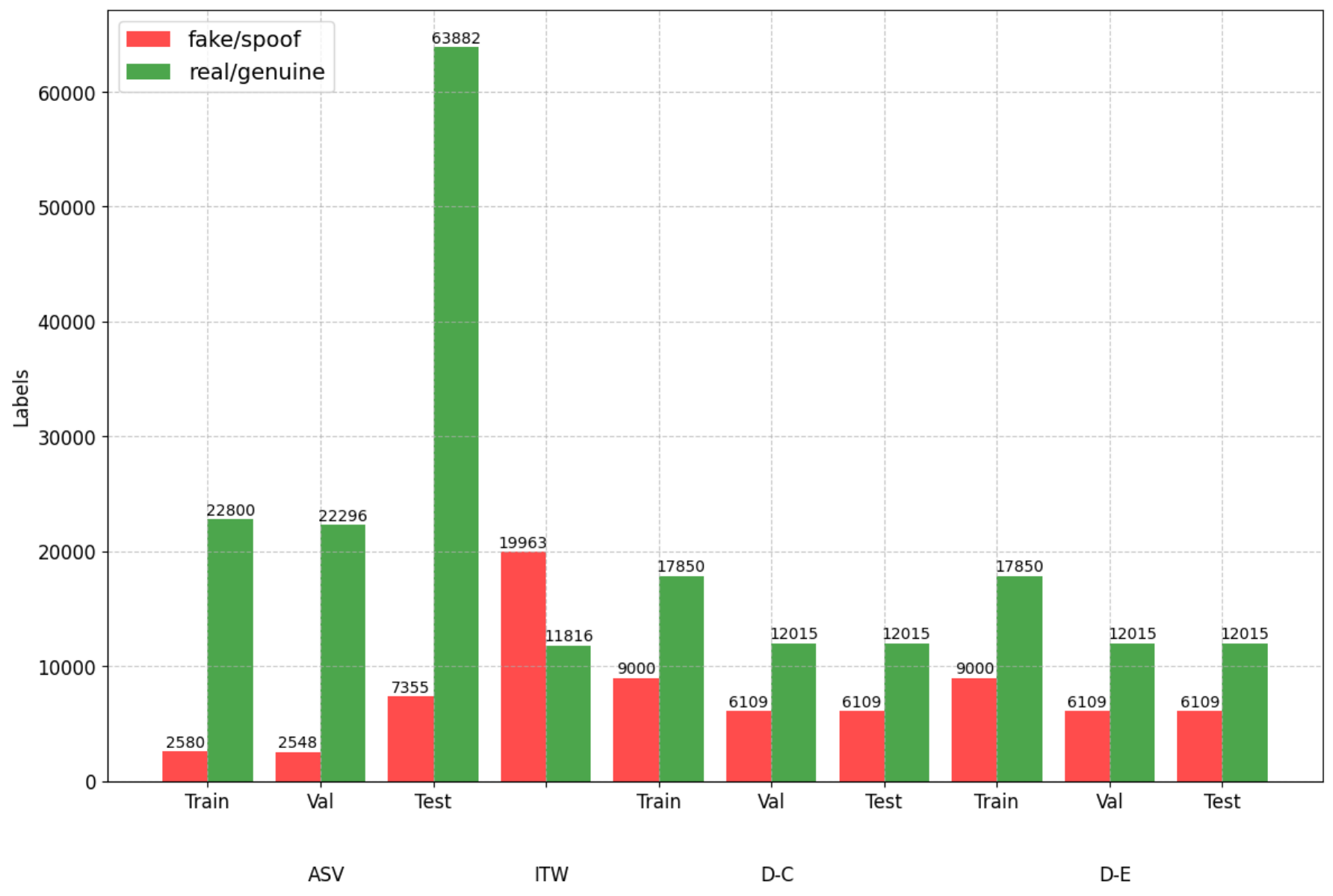}
    \caption{Class-wise data distribution across the datasets; D-C and D-E represents DECRO Chinese and English Set}
    \label{Datasets Label}
\end{figure}

The distribution of the audio clips with real and fake labels for each database is shown in Figure \ref{Datasets Label}.

\subsection{Detailed Information of the Pre-Trained Models}
\label{sec:Pre-trained Model Embedding}
Here, we describe various PTMs considered for our work. They are as follows:\newline 
\noindent \textbf{XLS-R}: It is multilingual representation learning model based on Wav2vec2 architecture. Training is carried out in a self-supervised manner and the objective involves solving a contrastive task over masked feature encoder outputs. It is trained on 436k hours of speech data comprising corpora VoxPopuli, Multilingual Librispeech, CommonVoice, VoxLingua107, and BABEL. XLS-R improves over XLS-R-53 pre-trained on 53 languages for various downstream speech processing tasks. We use the base\footnote{\url{https://huggingface.co/facebook/wav2vec2-xls-r-1b}} version comprising of 1 billion parameters for our work. \par

\noindent \textbf{Whisper}: Pre-training is carried out on 680k hours of data, incorporating a multitask format in a weakly-supervised way. It is an encoder-decoder-based modeling architecture. It is trained with the primary purpose of predicting transcriptions of audio content available on the internet. Whisper shows improved performance on speech recognition over XLS-R. We exploit the base\footnote{\url{https://huggingface.co/openai/whisper-base}} version with 74 million parameters for our use case.\par

\noindent \textbf{Massively Multilingual Speech (MMS)}: It is pre-trained on over 500k hours of speech data. It is built upon the Wav2vec2 model architecture that consists of a convolutional encoder followed by a BERT-like transformer block. MMS also follows contrastive pre-training as Wav2vec2. Pre-training data consists of various datasets such as MMS-lab, FLEURS, BABEL, etc. We use the 1 billion parameters version\footnote{\url{https://huggingface.co/facebook/mms-1b}} openly available.

\noindent\textbf{Unispeech-SAT}: It utilizes a contrastive loss model in conjunction with multitask learning. During its pre-training, UniSpeech-SAT follows a speaker-aware format. It is pre-trained on 960 hours of Librispeech English speech data. We make use of the base\footnote{\url{https://huggingface.co/microsoft/unispeech-sat-base}} version consisting of 94.68 million parameters. \par

\noindent\textbf{WavLM}: In its pre-training phase, WavLM simultaneously learns to predict masked speech and perform denoising. This dual process equips WavLM to effectively handle complex aspects of speech data, including speaker identity and spoken content, among others. WavLM (base)\footnote{\url{https://huggingface.co/microsoft/wavlm-base}} and WavLM (Large)\footnote{\url{https://huggingface.co/microsoft/wavlm-large}} versions are used for our experiments with 94.70 million and 316.62 million parameters respectively. WavLM (Base) and WavLM (Large) were pre-trained on 960 hours of Librispeech English data and Mix 94k data respectively. \par 
\noindent\textbf{Wav2vec2}: During training, Wav2vec2 masks the speech input in the latent space and completes a contrastive task defined over a quantization of the jointly learned latent representations. It was pre-trained on 960 hours of Librispeech English data. We choose the base\footnote{\url{https://huggingface.co/facebook/wav2vec2-base}} version with 95.04 million parameters and containing 12 transformer encoder blocks. \par

\noindent\textbf{x-vector\footnote{\url{https://huggingface.co/speechbrain/spkrec-xvect-voxceleb}}}: We took x-vector from \textit{Speechbrain} \cite{speechbrain} library. x-vector is a time-delay neural network trained in a supervised fashion for speaker recognition and achieves higher performance in comparison with the previous SOTA speaker recognition system, i-vector. It is trained on the combination of training data of Voxceleb1 and VoxCeleb2 with approx 4.2 million parameters. \par

 \noindent\textbf{XLSR\_emo\footnote{\url{https://huggingface.co/CAiRE/SER-wav2vec2-large-xlsr-53-eng-zho-all-age}}}: It is an XLS-R-53 \cite{conneau21_interspeech} model fine-tuned on training sets of various English and Chinese speech-emotion recognition databases such as CREMA-D, CSED, ElderReact, ESD, IEMOCAP, and TESS. \par

The input audio is sampled to 16KHz before passing as input to the PTMs. We extract the last hidden states from XLS-R, MMS, Unispeech-SAT, WavLM (Base), WavLM (Large), Wav2vec2, XLSR\_emo and convert the hidden states to vectors of dimensions 1280, 1280, 768, 768, 1024, 768, and 1024, respectively through the application of mean pooling. We discard the decoder for Whisper and extract the hidden representations from the encoder and through average pooling, we convert the representations to a vector of 512-dimension. Similarly, for the x-vector, we extract representations as vector size of 512-dimension.

\begin{figure*}[htbp]
    \begin{subfigure}{0.2\textwidth}
        \centering
        \includegraphics[width=\linewidth]{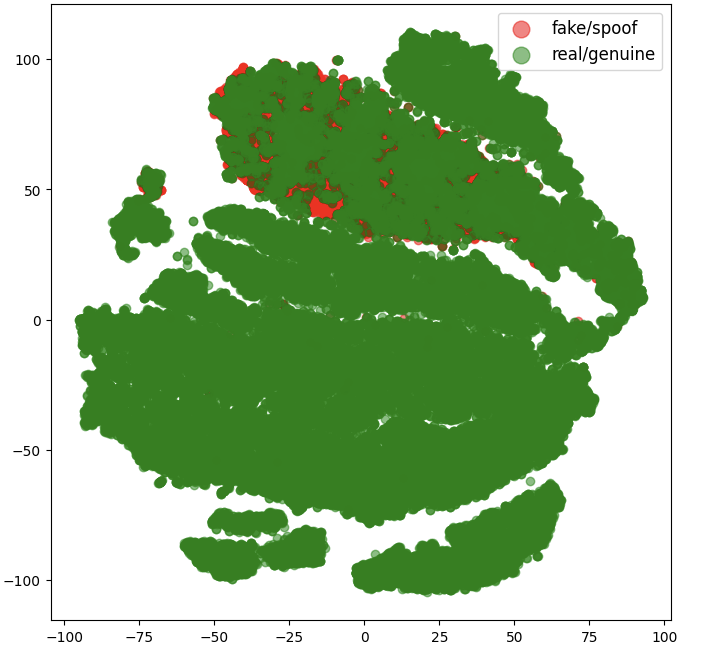}
        \caption{XLS-R}
    \end{subfigure}%
    \begin{subfigure}{0.19\textwidth}
        \centering
        \includegraphics[width=\linewidth]{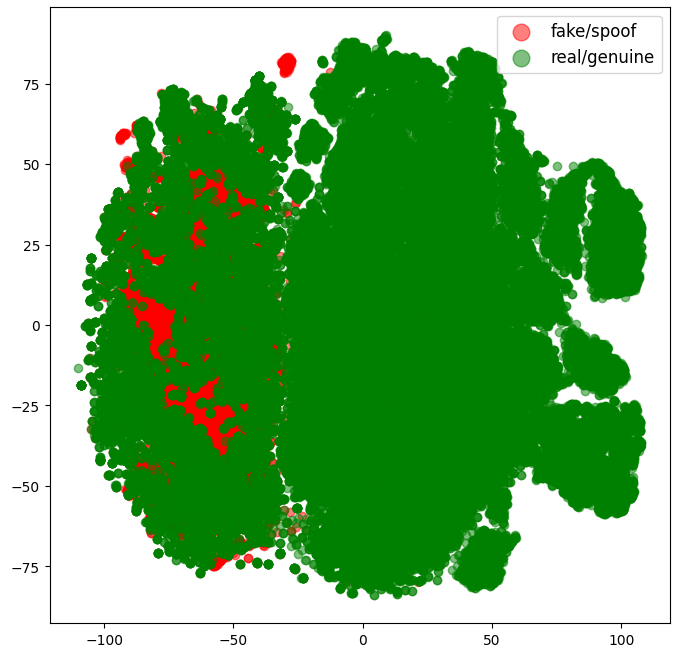}
        \caption{Whisper}
    \end{subfigure}%
    \begin{subfigure}{0.20\textwidth}
        \centering
        \includegraphics[width=\linewidth]{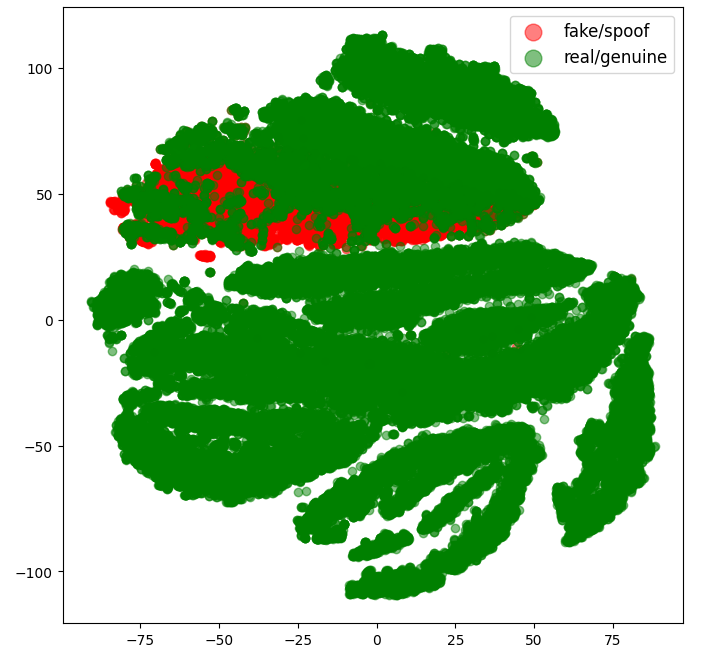}
        \caption{MMS}
    \end{subfigure}%
    \begin{subfigure}{0.201\textwidth}
        \centering
        \includegraphics[width=\linewidth]{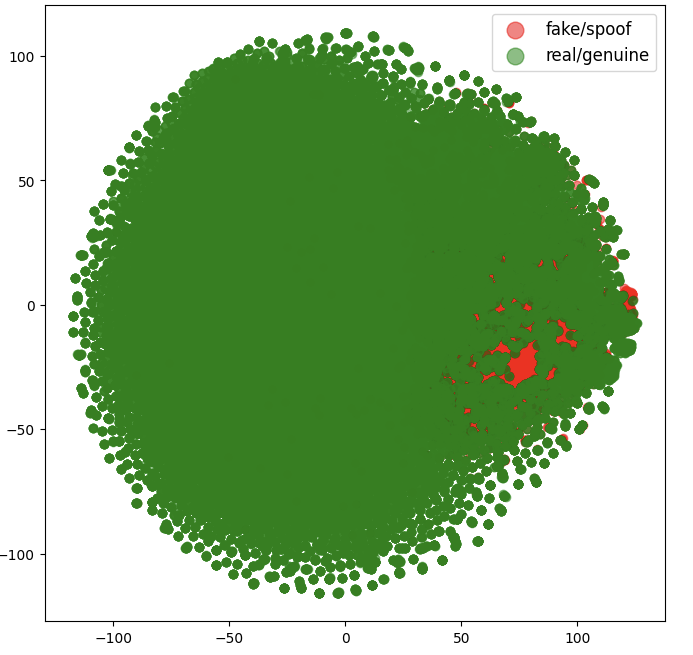}
        \caption{Unispeech-SAT}
    \end{subfigure}%
    \begin{subfigure}{0.2\textwidth}
        \centering
        \includegraphics[width=\linewidth]{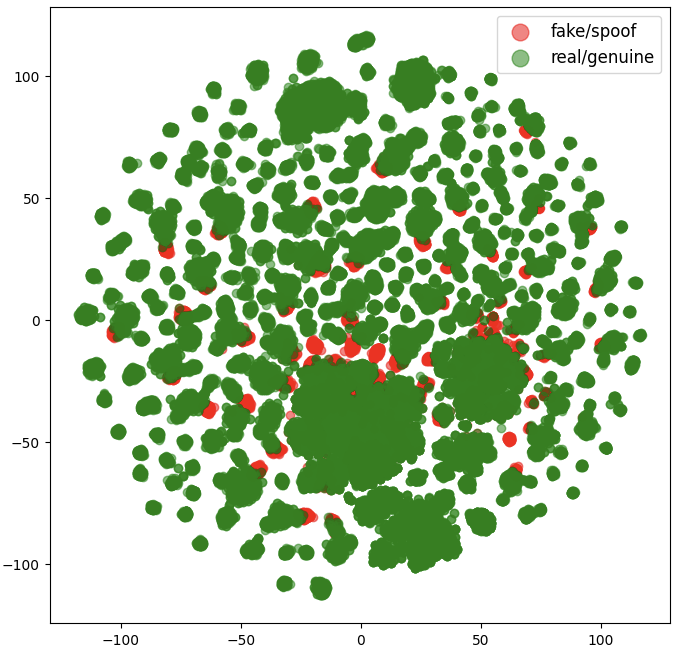}
        \caption{x-vector}
    \end{subfigure}%
    \caption{t-SNE plots of different PTM representations on ASV}
    \label{tsneasv}
\end{figure*}

\begin{figure*}[htbp]
    \begin{subfigure}{0.20\textwidth}
        \centering
        \includegraphics[width=\linewidth]{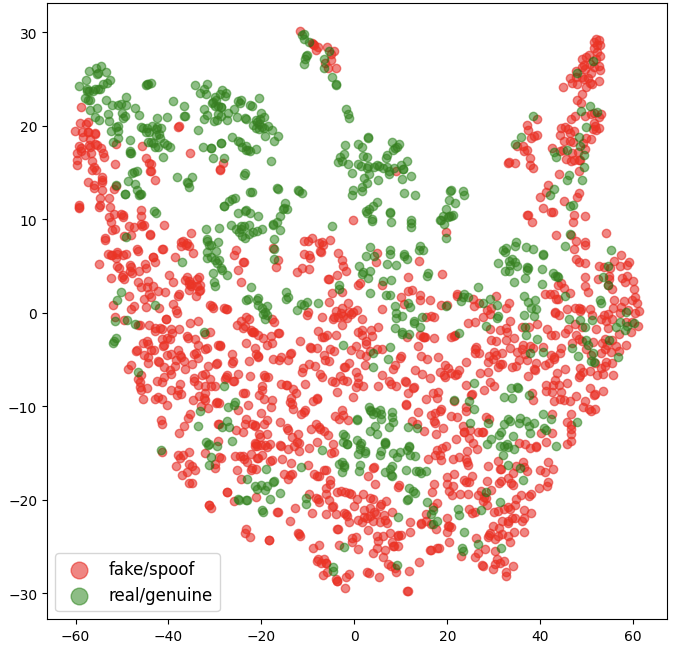}
        \caption{XLS-R}
    \end{subfigure}%
    \begin{subfigure}{0.205\textwidth}
        \centering
        \includegraphics[width=\linewidth]{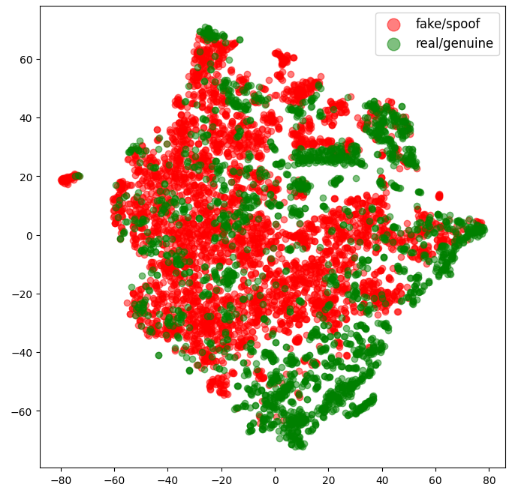}
        \caption{Whisper}
    \end{subfigure}%
    \begin{subfigure}{0.20\textwidth}
        \centering
        \includegraphics[width=\linewidth]{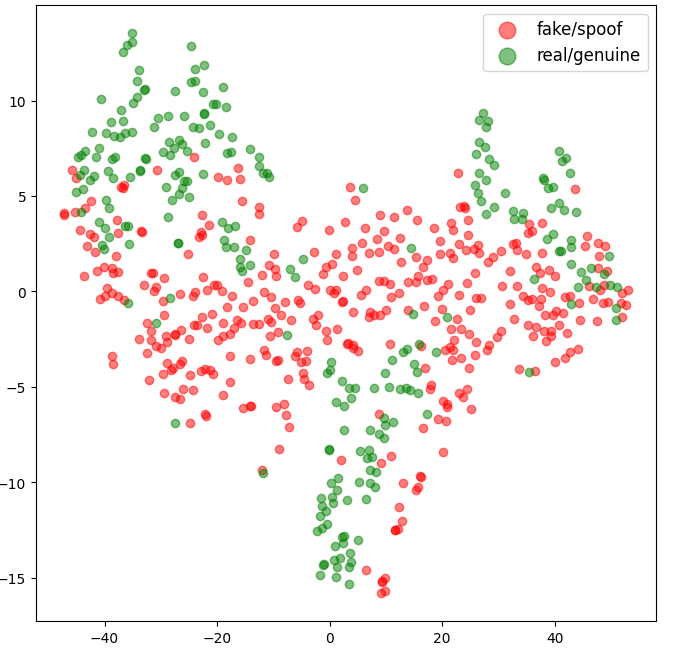}
        \caption{MMS}
    \end{subfigure}%
    \begin{subfigure}{0.2\textwidth}
        \centering
        \includegraphics[width=\linewidth]{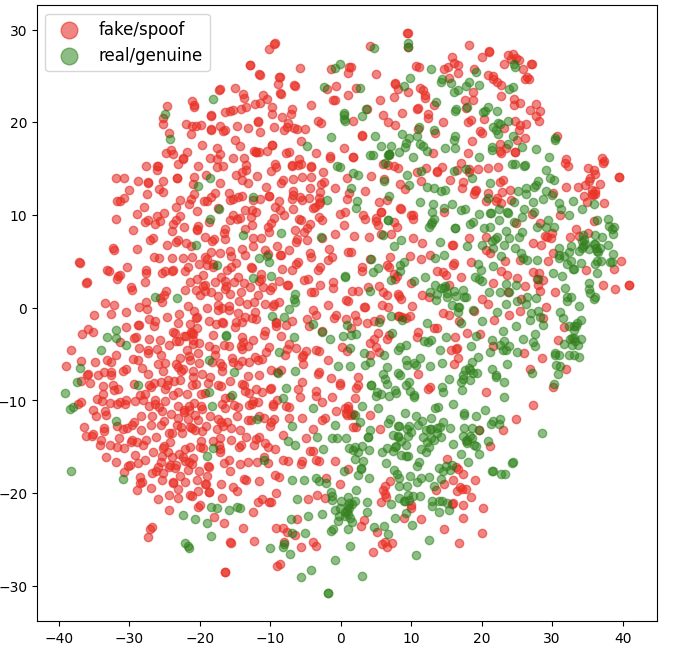}
        \caption{Unispeech-SAT}
    \end{subfigure}%
    \begin{subfigure}{0.21\textwidth}
        \centering
        \includegraphics[width=\linewidth]{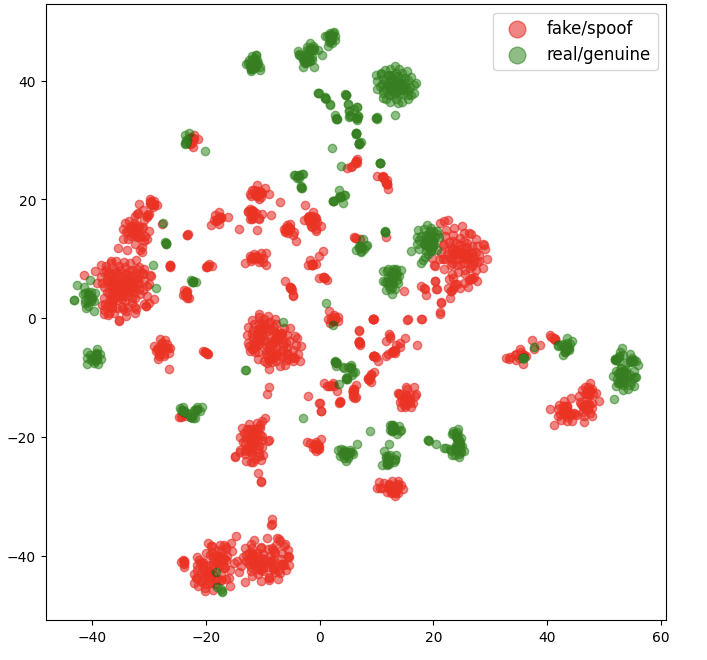}
        \caption{x-vector}
    \end{subfigure}%
    \caption{t-SNE plots of different PTM representations on ITW}
    \label{tsneitw}
\end{figure*}

\begin{figure*}[htbp]
    \begin{subfigure}{0.20\textwidth}
        \centering
        \includegraphics[width=\linewidth]{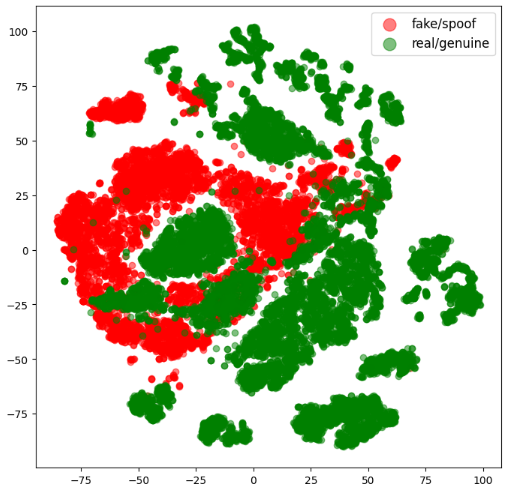}
        \caption{XLS-R}
    \end{subfigure}%
    \begin{subfigure}{0.2\textwidth}
        \centering
        \includegraphics[width=\linewidth]{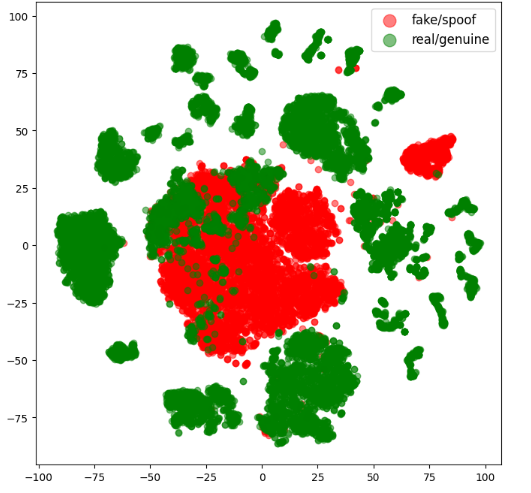}
        \caption{Whisper}
    \end{subfigure}%
    \begin{subfigure}{0.2\textwidth}
        \centering
        \includegraphics[width=\linewidth]{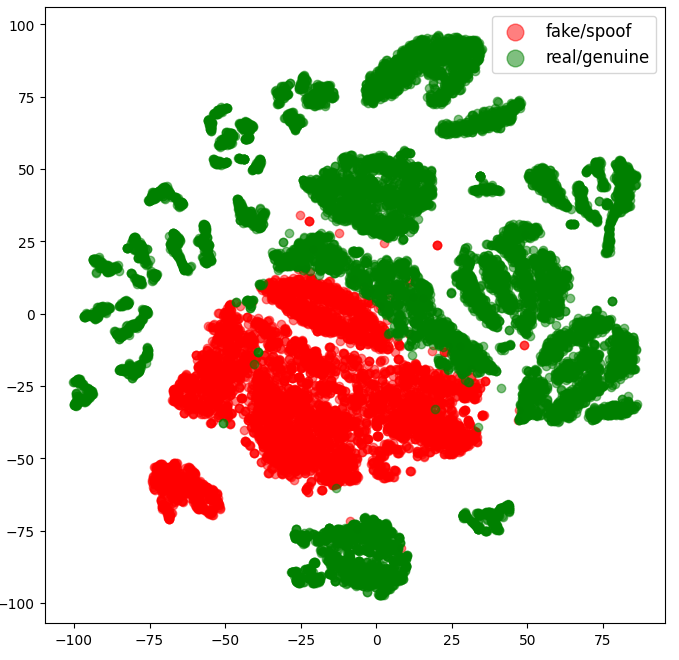}
        \caption{MMS}
    \end{subfigure}%
    \begin{subfigure}{0.2\textwidth}
        \centering
        \includegraphics[width=\linewidth]{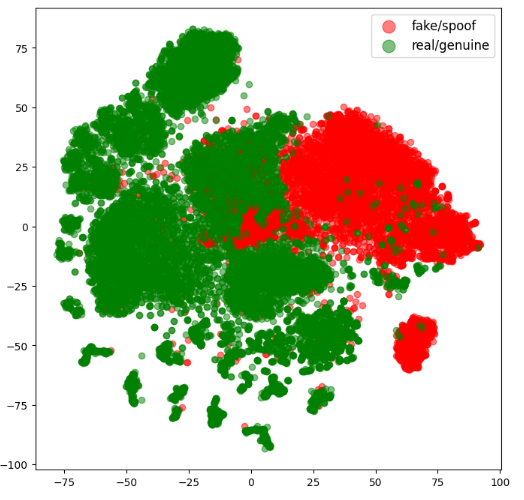}
        \caption{Unispeech-SAT}
    \end{subfigure}%
    \begin{subfigure}{0.208\textwidth}
        \centering
        \includegraphics[width=\linewidth]{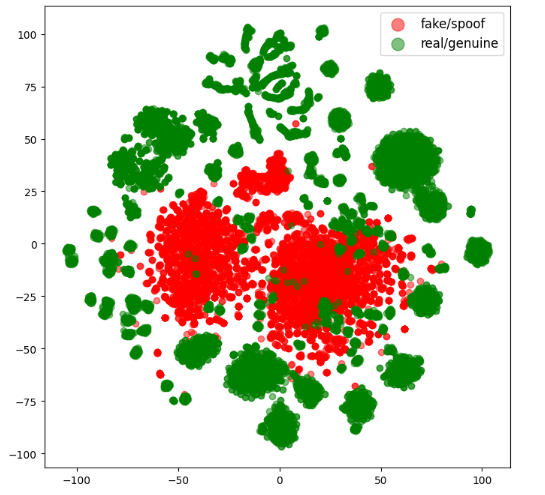}
        \caption{x-vector}
    \end{subfigure}%
    \caption{t-SNE plots of different PTM representations on D-C}
    \label{tsnedc}
\end{figure*}

\begin{table*}[h]
\scriptsize
\centering
\resizebox{\textwidth}{!}{
\begin{tabular}{c|c|c|c|c|c|c|c|c|c|c|c|c}
\hline
\multirow{2}{*}{\textbf{PTM}} & \multicolumn{3}{c|}{\textbf{ASV Training}} & \multicolumn{3}{c|}{\textbf{D-C Training}} & \multicolumn{3}{c|}{\textbf{ITW Training}} & \multicolumn{3}{c}{\textbf{D-E Training}} \\
\cline{2-13}
& \textbf{D-C} & \textbf{D-E} & \textbf{ITW} & \textbf{ASV} & \textbf{ITW} & \textbf{D-E} & \textbf{ASV} & \textbf{D-C} & \textbf{D-E} & \textbf{ASV} & \textbf{D-C} & \textbf{ITW} \\
\hline
XLS-R & \cellcolor{yellow!25} \textbf{22.78} & \cellcolor{green!25}\textbf{12.21}  & \cellcolor{yellow!25}\textbf{28.11} & \cellcolor{yellow!25}\textbf{10.23} & \cellcolor{yellow!25}\textbf{19.78} & \cellcolor{yellow!25}\textbf{15.32} & \cellcolor{blue!25}\textbf{10.45} & \cellcolor{blue!25}\textbf{18.88} & \cellcolor{yellow!25}\textbf{21.03} & \cellcolor{blue!25}\textbf{11.06} & \cellcolor{green!25}\textbf{39.58} & \cellcolor{blue!25}\textbf{14.94} \\
Whisper & \cellcolor{green!25} \textbf{30.51} & \cellcolor{yellow!25}\textbf{11.09}  & \cellcolor{green!25} \textbf{29.21} & \cellcolor{green!25}\textbf{17.19} & \cellcolor{green!25} \textbf{35.55} & \cellcolor{green!25} \textbf{15.34} & \cellcolor{yellow!25}\textbf{16.91} & \cellcolor{yellow!25}\textbf{39.62} & \cellcolor{green!25}\textbf{30.30} & \cellcolor{green!25}\textbf{15.78} & \cellcolor{blue!25}\textbf{30.67} & \cellcolor{yellow!25}\textbf{17.18} \\
MMS & \cellcolor{blue!25}\textbf{19.62} & \cellcolor{blue!25}\textbf{9.61} & \cellcolor{blue!25}\textbf{18.83} & \cellcolor{blue!25}\textbf{4.11} & \cellcolor{blue!25}\textbf{19.51} & \cellcolor{blue!25}\textbf{8.71} & 27.50 & 50.63 & \cellcolor{blue!25}\textbf{20.19} & \cellcolor{yellow!25} \textbf{14.78}  & \cellcolor{yellow!25}\textbf{32.75} & \cellcolor{green!25} \textbf{21.59}\\
Wav2Vec2 & 48.23 & 19.48  & 43.01 & 19.82 & 40.07 & 18.74 & 45.37 & 49.15 & 49.99 & 28.80 & 46.21 & 35.03 \\
WavLM (Large) & 31.10 & 13.19  & 32.12 & 31.23 & 36.69 & 15.99 & 28.13 & \cellcolor{green!25}\textbf{43.91} & 35.93 & 16.61 & 40.60 & 24.61 \\
x-vector & 32.98 & 13.62  & 37.43 & 39.53 & 38.66 & 17.69 & \cellcolor{green!25}\textbf{25.43} & 47.90 & 37.95 & 18.62 & 39.65 & 28.62 \\
\hline
\end{tabular}
}

\caption{Cross-Corpus Evaluation with representations of different PTMs kept at 120-dimension; Scores are in EER(\%); ASV Training, D-C Training, ITW Training, D-E Training represents training dataset and evaluated on the other datasets; For ITW, we select a test set for one splitting seed}
\label{120_cross}
\end{table*}

\begin{table*}[h]
\scriptsize
\centering
\resizebox{\textwidth}{!}{
\begin{tabular}{c|c|c|c|c|c|c|c|c|c|c|c|c}
\hline
\multirow{2}{*}{\textbf{PTM}} & \multicolumn{3}{c|}{\textbf{ASV Training}} & \multicolumn{3}{c|}{\textbf{D-C Training}} & \multicolumn{3}{c|}{\textbf{ITW Training}} & \multicolumn{3}{c}{\textbf{D-E Training}} \\
\cline{2-13}
& \textbf{D-C} & \textbf{D-E} & \textbf{ITW} & \textbf{ASV} & \textbf{ITW} & \textbf{D-E} & \textbf{ASV} & \textbf{D-C} & \textbf{D-E} & \textbf{ASV} & \textbf{D-C} & \textbf{ITW} \\
\hline
XLS-R & \cellcolor{blue!25}\textbf{21.33} & \cellcolor{blue!25}\textbf{12.78} & \cellcolor{yellow!25}\textbf{29.28} & \cellcolor{blue!25}\textbf{12.14} & \cellcolor{blue!25}\textbf{23.22} & \cellcolor{yellow!25}\textbf{21.88} & \cellcolor{blue!25}\textbf{8.21} & \cellcolor{blue!25}\textbf{17.69} & \cellcolor{green!25}\textbf{29.88} & \cellcolor{blue!25}\textbf{21.99} &  \cellcolor{yellow!25}\textbf{11.93} &  \cellcolor{blue!25}\textbf{14.65} \\
Whisper & \cellcolor{yellow!25}\textbf{25.22} & \cellcolor{yellow!25}\textbf{21.00} & \cellcolor{green!25}\textbf{34.88} & \cellcolor{yellow!25}\textbf{13.66} & \cellcolor{yellow!25}\textbf{29.22} & \cellcolor{blue!25}\textbf{16.94} & \cellcolor{yellow!25}\textbf{17.82} & \cellcolor{yellow!25}\textbf{33.60}  & \cellcolor{yellow!25}\textbf{17.87} & \cellcolor{green!25}\textbf{24.88}  & \cellcolor{blue!25}\textbf{11.87} & \cellcolor{green!25}\textbf{21.86}\\
MMS & \cellcolor{green!25}\textbf{34.61} & \cellcolor{green!25}\textbf{26.14} & \cellcolor{blue!25}\textbf{19.28} & \cellcolor{green!25}\textbf{19.88} & \cellcolor{green!25}\textbf{31.40} & \cellcolor{green!25}\textbf{32.97} & \cellcolor{green!25}\textbf{20.74} & 42.11 & \cellcolor{blue!25}\textbf{10.85} & \cellcolor{yellow!25}\textbf{23.16} & \cellcolor{green!25}\textbf{18.42} & \cellcolor{yellow!25}\textbf{21.43}\\
Wav2Vec2 & 57.66 & 43.98 & 46.66 & 44.63 &  47.22 & 41.90  & 27.34 & 53.33 &  44.97 & 35.76&  43.98& 46.95  \\
WavLM (Large) & 35.19 & 30.31 & 42.20  & 34.99 &  33.21&  34.28 & 21.11 & \cellcolor{green!25}\textbf{41.39} & 30.83 & 29.51 & 33.47 & 35.11 \\
x-vector & 35.79 & 32.34 & 45.04  & 37.09 &  39.22&  39.02 & 21.12 & 43.22& 31.00 & 32.55 & 32.77 & 32.98 \\
\hline
\end{tabular}
}
\caption{Cross-Corpus Evaluation with representations of different PTMs kept at 240-dimension; Scores are in EER(\%); ASV Training, D-C Training, ITW Training, D-E Training represents training dataset and evaluated on the other datasets; For ITW, we select a test set for one splitting seed}
\label{240_cross}
\end{table*}

\subsection{Cross-Corpus Evaluation}

We also investigate the cross-corpus generalization capability of the models trained on multilingual PTM representations as it has been shown in the literature of ADD \cite{ba2023transferring, muller22_interspeech} that models trained in one dataset or a certain language fail to perform in others. We use the same modeling approach as Figure \ref{1dcnn}. As the representations from different PTMs are of different dimensions, we use Principal Component Analysis (PCA) to transform the representations to the same dimension. We set the final dimension size after PCA to 120 and 240. We train the models on one training set of one dataset and evaluate on the test set of the others. We keep the training details like number of epochs, batch size, etc same as in Section \ref{Downstream Model}. We compare the multilingual PTMs with a monolingual PTM (Wav2vec2) and also speaker recognition PTM (x-vector) which reported competitive results (Table \ref{cnnsingle}) for better understanding of their generalization abilities. \par

The results of our experiments are presented in Table \ref{120_cross} and \ref{240_cross}. Models trained with multilingual PTM representations performed the best and this shows their cross-corpus generalization abilities.  However, the multilingual PTMs shows fluctuating performance among them, in some instances, representations from MMS performed the best, such as in Table  
\ref{120_cross} when trained on ASV and D-C Training set, whereas we achieve competitive results with XLS-R when trained on ITW and D-E and tested on the others. We notice significant differences in the results obtained across 120 and 240-dimension sizes. This points out that the dimension size of the representations also plays a minor role in the performance achieved in the downstream task. We also present cross-corpus evaluation scores for selected representations pairs with \textbf{MiO} in Table \ref{cc120} and \ref{cc240}.

\begin{table*}[h]
\scriptsize
\centering
\resizebox{\textwidth}{!}{
\begin{tabular}{c|c|c|c|c|c|c|c|c|c|c|c|c}
\hline
\multirow{2}{*}{\textbf{Model}} & \multicolumn{3}{c|}{\textbf{ASV Training}} & \multicolumn{3}{c|}{\textbf{D-C Training}} & \multicolumn{3}{c|}{\textbf{ITW Training}} & \multicolumn{3}{c}{\textbf{D-E Training}} \\
\cline{2-13}
& \textbf{D-C} & \textbf{D-E} & \textbf{ITW} & \textbf{ASV} & \textbf{ITW} & \textbf{D-E} & \textbf{ASV} & \textbf{D-C} & \textbf{D-E} & \textbf{ASV} & \textbf{D-C} & \textbf{ITW} \\
\hline
XLS-R + x-vector & \cellcolor{blue!25}\textbf{21.11} & \cellcolor{yellow!25}\textbf{12.04} & \cellcolor{blue!25}\textbf{24.80} & \cellcolor{blue!25}\textbf{16.53} & \cellcolor{blue!25}\textbf{23.34} & \cellcolor{yellow!25}\textbf{16.14} & \cellcolor{yellow!25}\textbf{58.78} & \cellcolor{yellow!25}\textbf{49.84} & \cellcolor{yellow!25}\textbf{69.14} & \cellcolor{yellow!25}\textbf{21.34} & \cellcolor{blue!25}\textbf{35.72} & \cellcolor{yellow!25}\textbf{48.15} \\

Whisper + Unispeech-SAT & \cellcolor{yellow!25}\textbf{44.59} & \cellcolor{blue!25}\textbf{7.80} & \cellcolor{yellow!25}\textbf{38.86} & \cellcolor{yellow!25}\textbf{26.89} & \cellcolor{yellow!25}\textbf{47.02} & \cellcolor{blue!25}\textbf{15.74} & \cellcolor{blue!25}\textbf{55.02} & \cellcolor{blue!25}\textbf{29.94} & \cellcolor{blue!25}\textbf{47.72} & \cellcolor{blue!25}\textbf{12.89} & \cellcolor{yellow!25}\textbf{44.26} & \cellcolor{blue!25}\textbf{27.50} \\
\hline
\end{tabular}
}
\caption{Cross-Corpus Evaluation with combined representations kept at 120-dimension; Scores are in EER(\%); ASV Training, D-C Training, ITW Training, D-E Training represents training dataset and evaluated on the other datasets; For ITW, we select a test set for one splitting seed}
\label{cc120}
\end{table*}

\begin{table*}[h]
\scriptsize
\centering
\resizebox{\textwidth}{!}{
\begin{tabular}{c|c|c|c|c|c|c|c|c|c|c|c|c}
\hline
\multirow{2}{*}{\textbf{Model}} & \multicolumn{3}{c|}{\textbf{ASV Training}} & \multicolumn{3}{c|}{\textbf{D-C Training}} & \multicolumn{3}{c|}{\textbf{ITW Training}} & \multicolumn{3}{c}{\textbf{D-E Training}} \\
\cline{2-13}
& \textbf{D-C} & \textbf{D-E} & \textbf{ITW} & \textbf{ASV} & \textbf{ITW} & \textbf{D-E} & \textbf{ASV} & \textbf{D-C} & \textbf{D-E} & \textbf{ASV} & \textbf{D-C} & \textbf{ITW} \\
\hline
XLS-R + x-vector & \cellcolor{blue!25}\textbf{17.21} & \cellcolor{blue!25}\textbf{15.14} & \cellcolor{blue!25}\textbf{24.53} & \cellcolor{blue!25}\textbf{14.84} & \cellcolor{blue!25}\textbf{17.97} & \cellcolor{blue!25}\textbf{13.53} & \cellcolor{yellow!25}\textbf{55.85} & \cellcolor{yellow!25}\textbf{50.91} & \cellcolor{yellow!25}\textbf{57.04} & \cellcolor{blue!25}\textbf{37.68} & \cellcolor{blue!25}\textbf{14.23} & \cellcolor{blue!25}\textbf{26.91} \\

Whisper + Unispeech-SAT & \cellcolor{yellow!25}\textbf{31.21} & \cellcolor{yellow!25}\textbf{15.70} & \cellcolor{yellow!25}\textbf{41.56} & \cellcolor{yellow!25}\textbf{24.61} & \cellcolor{yellow!25}\textbf{46.94} & \cellcolor{yellow!25}\textbf{18.52} & \cellcolor{blue!25}\textbf{54.80} & \cellcolor{blue!25}\textbf{39.57} & \cellcolor{blue!25}\textbf{46.75} & \cellcolor{yellow!25}\textbf{42.57} & \cellcolor{yellow!25}\textbf{18.34} & \cellcolor{yellow!25}\textbf{31.09} \\
\hline
\end{tabular}
}
\caption{Cross-Corpus Evaluation with combined representations kept at 240-dimension; Scores are in EER(\%); ASV Training, D-C Training, ITW Training, D-E Training represents training dataset and evaluated on the other datasets; For ITW, we select a test set for one splitting seed}
\label{cc240}
\end{table*}

\end{document}